\documentclass[a4paper,11pt, twocolumn]{article}

\usepackage{graphicx}
\usepackage{grffile}
\usepackage{multicol}
\usepackage{amsmath}
\usepackage{amssymb}
\usepackage{bm}
\usepackage{wrapfig}
\usepackage{pdfpages}

\usepackage{color}
\usepackage{xcolor}
\usepackage{hyperref}

\hypersetup{
    colorlinks=true,
    allcolors=blue
    }

\usepackage[backend=biber]{biblatex}

\DefineBibliographyStrings{english}{andothers={+}}

\newbibmacro{name:newformat}{%
   \namepartfamily  % #1->\namepartfamily, #2->\namepartfamilyi
   }

\DeclareNameFormat{default}{%
  \nameparts{#1}% split the name data, will not be necessary in future versions
  \usebibmacro{name:newformat}%
  \usebibmacro{name:andothers}%
}

\newcounter{project}

\usepackage[format=plain,font=small,labelfont=bf,skip=-5pt,belowskip=-20pt]{caption}

\newcommand{\heading}[1]
{\par\vspace{2mm}\noindent{\large\fontseries{b}\selectfont{\bf #1}}}

\newcommand{\Condensed}{}
\newcommand{\sstrong}[1]{\textbf{#1}}

\newcommand{\subheading}[1]
{\par\vspace{1.5mm}\noindent{\Condensed\sstrong{#1}}}

\newcommand{\metadata}[1]
 {{\centering\small{#1}\par}\vspace{1mm}\ignorespaces}

\newenvironment{psummary}[1][]{\vspace{1mm}\par\hrule\vspace{0.5mm}\noindent\metadata{{#1}}{\par\noindent\sstrong{Summary\ }}}{\par\vspace{2mm}\hrule}

	{\end{list}}

\def\spose#1{\hbox to 0pt{#1\hss}}
\def\simlt{\mathrel{\spose{\lower 3pt\hbox{$\mathchar"218$}}
     \raise 2.0pt\hbox{$\mathchar"13C$}}}
\def\simgt{\mathrel{\spose{\lower 3pt\hbox{$\mathchar"218$}}
     \raise 2.0pt\hbox{$\mathchar"13E$}}}

\def\gtsima{$\; \buildrel > \over \sim \;$}
\def\ltsima{$\; \buildrel < \over \sim \;$}
\def\gsim{\lower.5ex\hbox{\gtsima}}
\def\lsim{\lower.5ex\hbox{\ltsima}}
\def\simgt{\lower.5ex\hbox{\gtsima}}
\def\simlt{\lower.5ex\hbox{\ltsima}}
\def\simpr{\lower.5ex\hbox{\prosima}}

\def\spose#1{\hbox to 0pt{#1\hss}}

\def\simgt{\mathrel{\spose{\lower 3pt\hbox{$\mathchar"218$}}
     \raise 2.0pt\hbox{$\mathchar"13E$}}}

\usepackage{lipsum}
\DeclareMathOperator{\cow}{\mathrm{cow}}

\begin{document}
 
%%%%%%%%%%%%%%%%%%%%%%%%%%%%%%%%%%%%%%%%%%%%%%%%%%%%%%%

\clearpage

% PROJECTS
\newcommand{\projectpath}{} % Necessary because of the inclusion of extra files
\begin{refsection} 
\twocolumn[
\title{\bf COWS all tHE way Down (COWSHED) I: \\ Could cow based planetoids support methane atmospheres?}
\author{
William J. Roper$^{1}$,
Todd L. Cook$^{1}$,
Violetta Korbina$^{1}$,
Jussi K. Kuusisto$^{1}$,
Roisin O'Connor$^{1}$,\\
Stephen D. Riggs$^{1}$,
David J. Turner$^{1}$,
Reese Wilkinson$^{1}$
}
\date{April 1st 2022}
% List of institutions
\maketitle
$^{1}$Astronomy Centre, University of Sussex, Falmer, Brighton BN1 9QH, UK
\begin{psummary}
More often than not a lunch time conversation will veer off into bizarre and uncharted territories. In rare instances these frontiers of conversation can lead to deep insights about the Universe we inhabit. This paper details the fruits of one such conversation. In this paper we will answer the question: How many cows do you need to form a planetoid entirely comprised of cows, which will support a methane atmoosphere produced by the planetary herd? We will not only present the necessary assumptions and theory underpinning the cow-culations, but also present a thorough (and rather robust) discussion of the viability of, and implications for accomplishing, such a feat.
\end{psummary}
]

% The introduction
\heading{Introduction}

Regarding cows, majestic beasts that have provided humanity with dairy, meat and lawn mowing for countless generations. As humanity has evolved we have learnt to cultivate these beasts and some cultures even revere them beyond a simple source of sustenance. However, their environmental impact cannot be ignored. The methane they produce as waste is an extremely damaging greenhouse gas. In this paper we aim to use this damaging property and ask the question: How many cows do you need to form a planetoid entirely comprised of cows which will support a methane atmosphere produced by the planetary herd?

\subheading{Assumptions}

For simplicity in this early work it is necessary to make several assumptions. Although, we note that some of these can certainly be relaxed at a later date.

\begin{enumerate}
    \item Cows are incompressible\footnote{Anyone that has had the pleasure of petting a cow can certainly attest to the validity of this assumption.}.
    \item The planetoid/herd is perfectly fed and hydrated. Of course, maintaining extra-terrestrial cattle is a significant challenge to this endeavour, but we leave this consideration to future work on the matter.
    \item The methane atmosphere is assumed to have the same average density as Earth's own atmosphere.
    \item Contrary to what many will have you believe, cows are not spherical. 
    \item Cows are sufficiently intelligent to organise themselves and obtain a perfect bovine packing fraction of unity, such that the volume of the planetoid is simply $V_{P}=N_{\cow}V_{\cow}$.
    \item The cows have negligible moomentum.
    \item Cows can be approximated as a blackbody with effective temperature $T_{\cow}$. 
\end{enumerate}

\heading{Cow-culations} 

To address this question we need an expression for the flux of methane leaving the planetoid. There are numerous constituent parts to this calculation, here we will present the important steps but leave the nitty gritty as an exercise for the reader.

The first port of call in this mathematical stroll towards bovine enlightenment is the escape velocity of methane. Starting with the expression for escape velocity,
\begin{equation}
v_{e}=\sqrt{\frac{2GM_{P}}{R_P + H}},
\end{equation}
where $H$ is the scale height of the atmosphere and the mass of a purely cow based planetoid can be described as
\begin{equation}
   M_{P}=N_{\cow}M_{\cow},
\end{equation}
assuming an average cow mass of $M_{\cow}=1390$ pounds $=630$ kg \cite{cowMass}. Using the expression above for the volume of the planet, and taking the volume of a cow to be $V_{\cow}=1.1$ m$^{3}$ \cite{cowVol}, we can derive the radius of the planetoid and arrive at the final expression for escape velocity as a function of only constants and cow properties,
\begin{equation}
    v_{e} = \sqrt{\frac{2GM_{\cow}N_{\cow}}{\left(\frac{3}{4}\frac{N_{\cow}V_{\cow}}{\pi}\right)^{1/3} + H}}.
\end{equation}

With the Newtonian mechanics out of the way we must now brave the world of thermal and statistical physics. To find the flux of methane lost by the planetoid we start with the Maxwell-Boltzmann distribution (MBD),
\begin{equation}
    f(v) dv= \left(\frac{m_g}{2\pi k T}\right)^{3/2}4\pi v^2 e^{-\left(\frac{m_g v^2}{2 k T}\right)}dv.
\end{equation}
We obtain the flux ($\Phi$) by performing some spherical coordinate gymnastics, multiplying by the velocity and gas density, and integrating the MBD over a hemisphere to capture the outward moving flux, yielding
\begin{equation}
    \Phi = \frac{\rho}{4}vf(v)dv.
\end{equation}
From this point the cow-culation devolves into triviality with the integration of the flux between the escape velocity and infinity giving the final flux expression
\begin{equation}
    \Phi=\alpha\left(\beta+\frac{2kT}{m_g}\right)e^{-\left(\frac{m_g}{kT}\beta\right)},
\end{equation}
where $m_g$ is the mass of a methane molecule, $T$ is the atmospheric temperature, $\alpha$ is given by 
\begin{equation}
    \alpha=\frac{\rho}{2^{3/2}}\left(\frac{m_g}{\pi k T}\right)^{1/2},
\end{equation}
and $\beta$ is a function of the number of cows forming the planetoid, singular cow mass and singular cow volume ($N_{\cow}$, $M_{\cow}$, and $V_{\cow}$),
\begin{equation}
    \beta=\frac{2GM_{\cow}N_{\cow}}{\left(\frac{3}{4}\frac{N_{\cow}V_{\cow}}{\pi}\right)^{1/3} + H}.
\end{equation}

With the meat of the work done, we are left with only two outstanding ingredients: the scale height of the atmosphere and the atmospheric temperature. For the scale height we employ the expression
\begin{equation}
    H=\frac{kT}{m_g g},
\end{equation}
where $g$ is the acceleration due to gravity on the planetoid. 

The atmospheric temperature is a little more involved, not only must we take into account the temperature induced by the flux from our Sun but we must also include the contribution due to the bovine blackbody temperature. This temperature is taken to be the maximum between $T_{\cow}=303.9$ K, where we have used the mean surface temperature taken from a study using infrared thermography \cite{cowTemp},
and the temperature of the planetoid's surface induced by the Sun,
\begin{equation}
    T_{\mathrm{surf}} = \left(\frac{L_\odot(1 - a_{\cow})}{16\pi\sigma D^2}\right)^{1/4},
\end{equation}
where $a_{\cow}=0.04$ is the albedo of a cow, assuming the worst case scenario of a black hide \cite{albedoCow}.

Combining the flux from the Sun, the temperature of the planetoid and the greenhouse effect, we arrive at the following expression for the atmospheric temperature,
\begin{equation}
    T^{4} = \frac{L_\odot(1 - a)}{16\pi\sigma D^2} + (1 - f)\times\mathrm{max}(T_{\cow}, T_{\mathrm{surf}})^4,
\end{equation} 
where $L_\odot$ is the Sun's luminosity, $a$ is the albedo of the atmosphere (for which we assume $a=0.3$), $D$ is the distance from the Sun, and $f$ is the fraction of energy absorbed by the atmosphere (for which we assume $f=0.8$). 

All that is left to do now is balance the flux of methane leaving the planetoid with the methane produced by the planetoid's bovine makeup. Due to the environmental impact of cattle farming there has been much research into the methane production of cattle and how this can be affected by their diet \cite[e.g.][]{methRate1, methRate2, methRate3, methRate4}. We utilise this wealth of data to derive a planetoid methane production rate of
\begin{equation}
    N_{\cow}\dot{M}_{\mathrm{CH}_{4}}=N_{\cow}(3.0\times 10^{-6} \ \mathrm{kg} / \mathrm{s}).
\end{equation}
Which can be converted into a flux by dividing by the surface area of the planetoid ($A_P$) yielding
\begin{equation}
    \Phi_{\cow} = \frac{N_{\cow}\dot{M}_{\mathrm{CH}_{4}}}{A_{P}}.
\end{equation}

In the final step we equate the two methane fluxes, equating loss and production, which leads to an analytically insoluble expression,
\begin{equation}
    \frac{N_{\cow}\dot{M}_{\mathrm{CH}_{4}}}{A_{P}}=\alpha\left(\beta+\frac{2kT}{m_g}\right)e^{-\left(\frac{m_g}{kT}\beta\right)}
\end{equation}
In the subsequent sections we present results from solving this expression numerically using the publicly available \texttt{Python} package \texttt{scipy} \cite{scipy}.

\heading{Cow Planet}

\begin{figure}[h]
    \centering
    \includegraphics[width=\columnwidth]{\projectpath 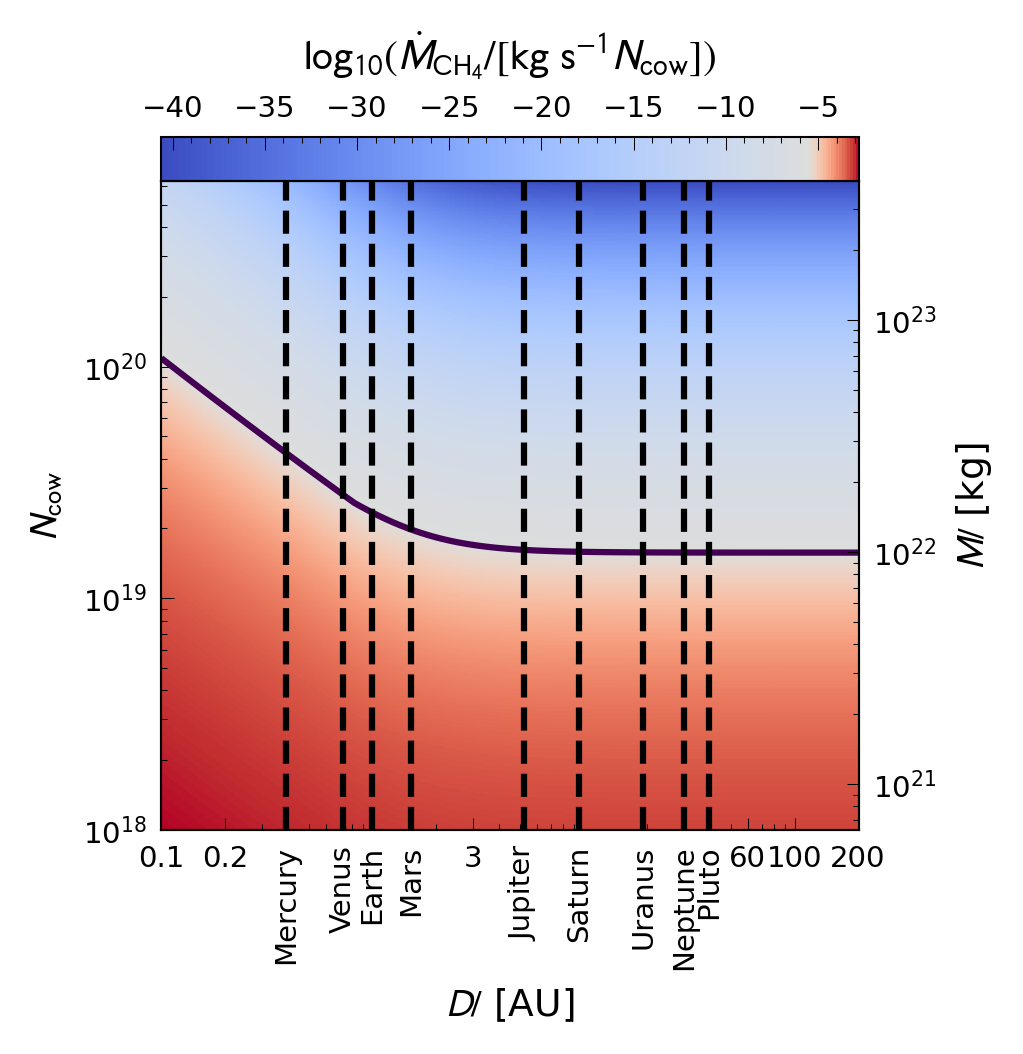}
    \vspace{0.1in}
    \caption{A plot showing the atmospheric loss and production as a function of distance from the Sun and the number of cows comprising the planetary herd (and therefore planetoid mass). The colormap indicates the state of the atmosphere; red indicates a regime where $\Phi>\Phi_{\cow}$ and an atmosphere cannot be supported, blue indicates a regime where $\Phi<\Phi_{\cow}$ and the atmosphere is growing as the cows production outweighs the methane loss. The curve represents the hyperplane in this parameter space where methane loss is balanced by methane production and a steady state atmosphere can be supported. The dashed vertical lines represent the distance of each planet in the solar system to aid interpretation.}
    \label{fig:cowatmos}
    \vspace{0.1in}
\end{figure}

With all the constituent pieces in place we can now move on to the analysis. In Fig. \ref{fig:cowatmos} we present the methane production as a function of the number of cows in the planetoid and the distance from the Sun. Beyond an orbital distance of $\sim3$ AU the atmospheric loss is driven by the bovine blackbody temperature. At this distance we reach a minimum of $\log_{10}(N_{\cow})\approx19.2$ required to balance the methane production and loss. Surprisingly, this means an atmosphere can be sustained at a mass of only $\sim0.1$\% the mass of the Earth with a bovine planetoid radius of $\sim24.9$\% the radius of Earth. However, it is worth taking into account that, unlike traditional atmospheres, this atmosphere is significantly replenished by the planetary body it is ``bound'' to; because of this fact the atmosphere is able to reach a steady state at a significantly lower mass than necessary for a celestial body without a similar atmosphere replenishment mechanism. Due to this low mass, we end up with an acceleration due to gravity of $g\approx0.3$ m s$^{-2}$ and thus a large atmospheric scale height of $H\approx410$ km.

\begin{figure}[h]
    \centering
    \includegraphics[width=\columnwidth]{\projectpath 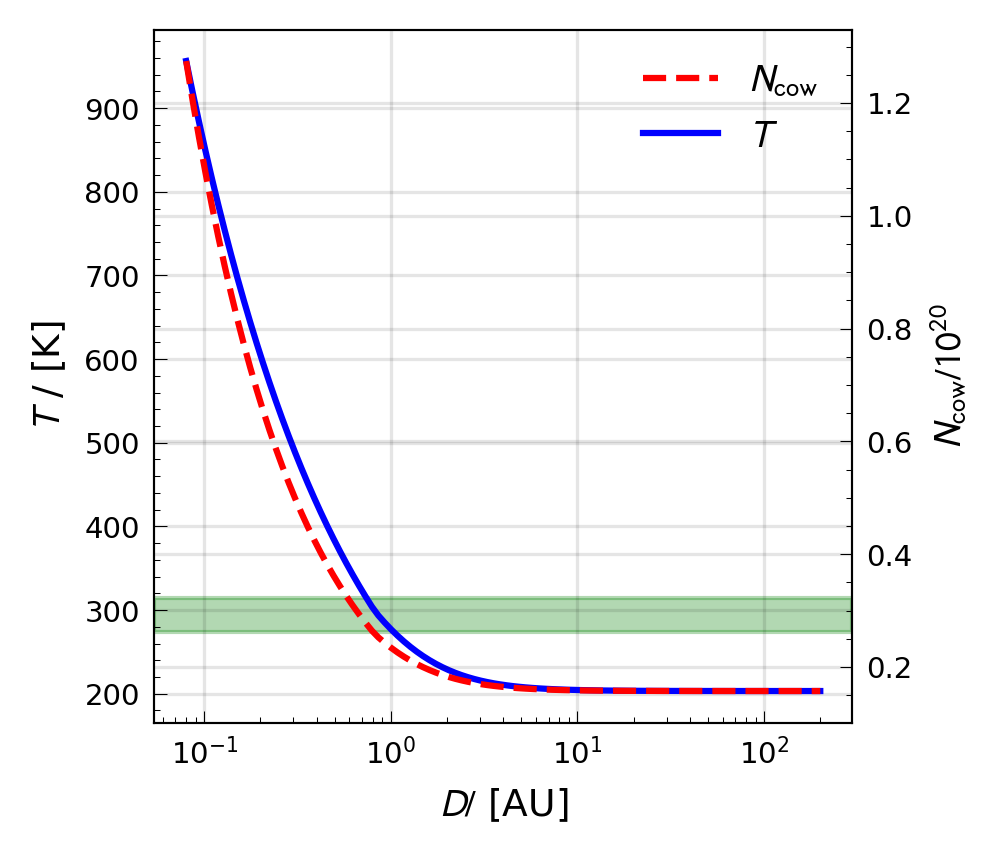}
    \vspace{0.1in}
    \caption{A plot showing the effects of distance from the Sun. The solid blue line shows how the atmospheric temperature scales, and the dashed red line shows how the number of cows scales for comparison. The shaded region shows the "habitable zone" for our cow planetoid where the atmosphere and cows themselves are in an acceptable temperature range for the cows to exist without further developments to bovine survivability.}
    \label{fig:atmostemp}
    \vspace{0.1in}
\end{figure}

As we shepherd our intrepid spacefaring bovines toward the Sun the flux from the Sun becomes a far more significant factor in the temperature of the atmosphere and we begin to require a larger herd. Although the increase in the number of cows required is actually quite modest with a planetoid at Mercury's orbit only requiring an increase of $\Delta\log_{10}(N_{\cow})\sim0.4$. This planetoid, despite a significant increase in atmospheric temperature, is capable of balancing the atmospheric loss, due to increasing temperatures, by sheer weight of methane production by our hardy herd.

In Fig. \ref{fig:atmostemp} we present the scaling of atmospheric temperature and number of cows for a steady state atmosphere as a function of distance from the Sun. Unsurprisingly the two scale similarly with distance, as the atmospheric temperature increases so does the methane loss, and this must be balanced by methane production (number of cows). However, cows evolved on Earth which means they are not well acclimatised to temperatures beyond the ranges found on Earth. Without further development of bovine survivability in extreme environments we must limit our hopes to the habitable zone shaded in green on Fig. \ref{fig:atmostemp}. Luckily for us, by definition, this habitable region is very close to Earth making the job of transporting the herd to their eventual location considerably easier. 

\heading{Cow Moon}

Of course, we could save a significant amount of bovine mass by employing existing celestial bodies as the starting point for our bovine planetoid. We note however, that this does pollute the purity of the bovine only celestial body. For this demonstration we will use the Moon as our testing ground.

The expressions presented above remain unchanged with the exception of planetary mass and radius which must now take into account the mass and radius of the Moon, and the temperature of the atmosphere which must now take into account the surface temperature of the Moon and the area of the Moon available to absorb flux from the Sun. Taking this into account we have a new expression for the atmospheric temperature given by

\begin{equation}
    T^{4} = T_{\mathrm{flux}}^{4} + (1 - f)\left(\mathrm{max}(T_{\cow}, T_{\mathrm{surf}})^4 + T_{\mathrm{Moon}}^4\right),
\end{equation} 
where
\begin{equation}
    T_{\mathrm{flux}}^4=\frac{L_\odot(1 - a)}{16\pi\sigma D^2},
\end{equation}
\begin{equation}
    T_{\mathrm{Moon}}^4=\frac{A_{\mathrm{abs}}}{A_{\mathrm{rad}}}\frac{L_\odot(1 - a)}{16\pi\sigma D^2},
\end{equation}
$A_{\mathrm{abs}}$ is the area of the Moon presented to the Sun not covered by cows, and $A_{\mathrm{rad}}$ is the area over which the energy is radiated. 

\begin{figure}[h]
    \centering
    \includegraphics[width=\columnwidth]{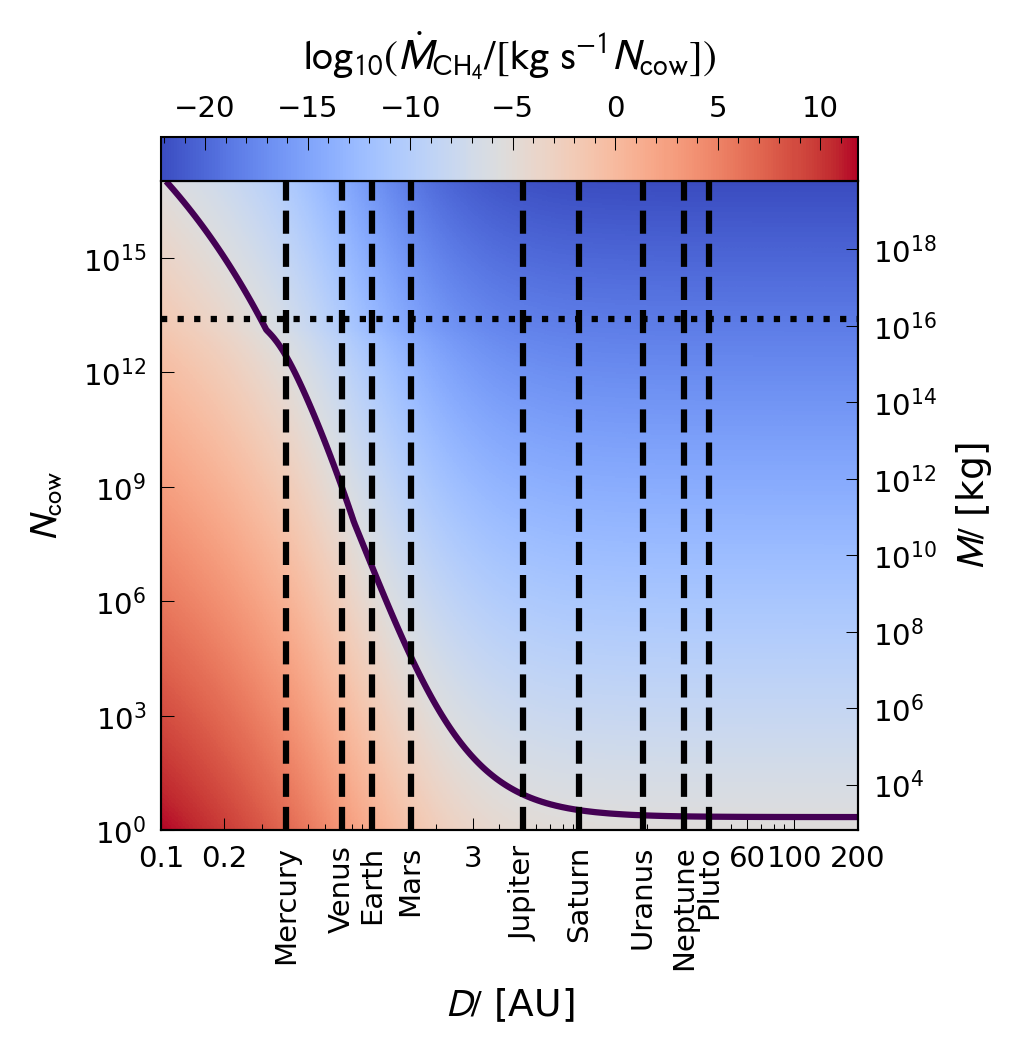}
    \vspace{0.1in}
    \caption{A plot showing the atmospheric loss and production as a function of distance from the Sun and the number of cows inhabiting the Moon (and therefore the mass of the bovine-lunar system). The colormap indicates the state of the atmosphere; red indicates a regime where $\Phi>\Phi_{\cow}$ and an atmosphere cannot be supported, blue indicates a regime where $\Phi<\Phi_{\cow}$ and the atmosphere is growing as the cows production outweighs the methane loss. The curve represents the hyperplane in this parameter space where methane loss is balanced by methane production and a steady state atmosphere can be supported. The dashed vertical lines represent the distance of each planet in the solar system to aid interpretation. The dotted horizontal line indicates the number cows at which the Moon's surface is completely obscured by a shell of cows. At this lunar herd size the Moon cannot absorb any flux from the Sun to contribute to the blackbody heating of the atmosphere.}
    \label{fig:cowatmos_Moon}
    \vspace{0.1in}
\end{figure}

Fig. \ref{fig:cowatmos_Moon} is a reproduction of Fig. \ref{fig:cowatmos} but taking into account the Moon as a starting point. Again it shows the production and loss of methane as a function of distance from the Sun, and lunar herd size and mass. We can now see a far larger variation in the number of cows in our lunar herd. In fact, rather than a bovine planetoid in this case for most orbits we simply have a population of lunar cows. Inside Mercury's orbit the number of cows is great enough to completely cover the surface of Moon and we once again have complete shells of cows surrounding the lunar core. Possibly most surprisingly, if our herd could withstand the temperatures at Jupiter's orbit and beyond, we would only need $<100$ cows to sustain a methane atmosphere on the Moon! 

Much like the bovine planetoid the fact that the atmosphere is constantly replenished and the comparatively small acceleration due to gravity leads to a larger scale height than found on Earth varying from $\sim140$ km at Mercury's orbit to $\sim64$ km at Jupiter's orbit and beyond.

\heading{Rearing extra-terrestrial cattle}

\begin{figure}[h]
    \centering
    \includegraphics[width=\columnwidth]{\projectpath 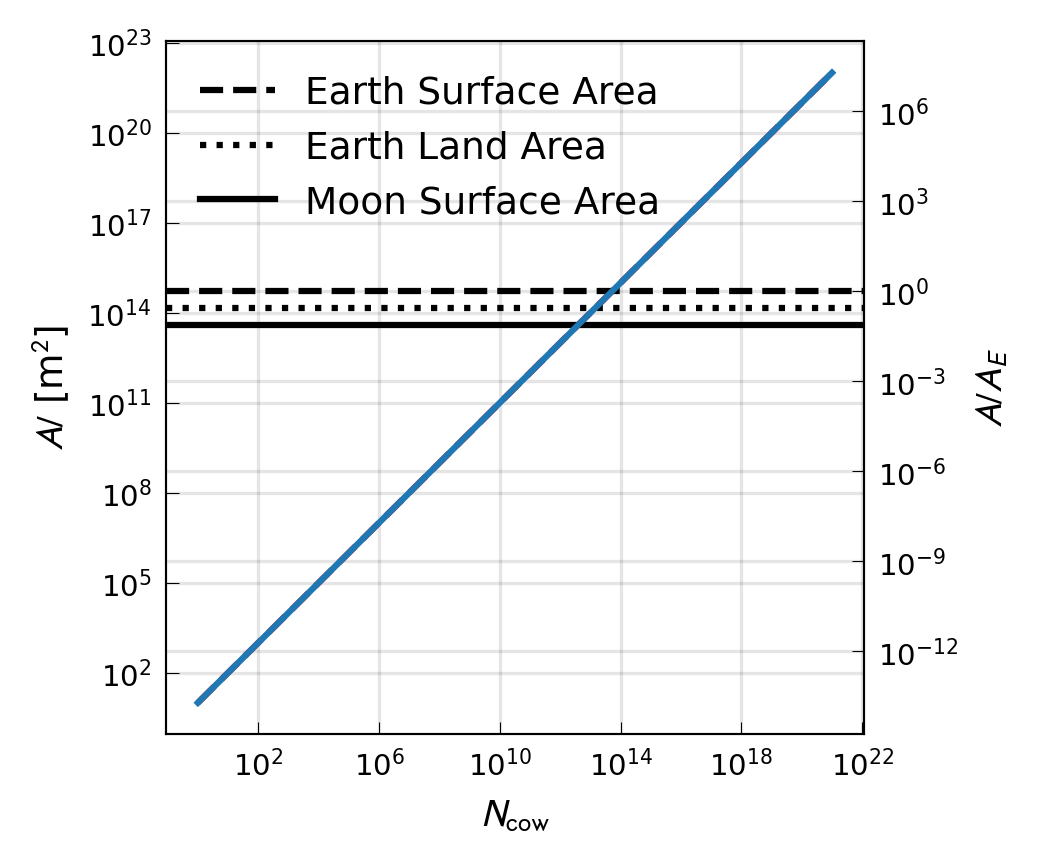}
    \vspace{0.1in}
    \caption{A plot demonstrating the area necessary to raise our spacefaring bovines prior to launch. The blue curve shows the area necessary for rearing cattle as recommended by the Agricultural and Horticultural DataBase (AHDB) \cite{cowFootprint}. The dotted line represents the area of land mass on Earth, the dashed line represents the Earth's entire surface area and the solid lines shows the surface area of the Moon.}
    \label{fig:area}
    \vspace{0.1in}
\end{figure}

This theory is all well and good, but how possible is this agricultural marvel given our current resources? Our cows require a significant amount of rearing before we can begin to consider them for launch. The first issue we must face is where we could do this rearing. Cows require approximately 10 m$^{2}$ of space according to recommendations from the Agricultural and Horticultural DataBase (AHDB) \cite{cowFootprint}, does this pose a problem? In Fig. \ref{fig:area} we plot the area necessary to care for our planetary herd, with some areas of import for comparison. Sadly things are looking bleak for the pure bovine planetoid, with a required number of $\sim10^{19}$ cows we couldn't support this herd even using the entire surface area Earth. Of course future advancements in agriculture may help us solve this issue, but for now this is but a pipe dream. More encouragingly, the $\sim10^6$ cows necessary to sustain an atmosphere on the Moon at Earth's radius (conveniently where the Moon resides) could be sustained using only a fraction of the land area of Earth. In fact, the best way forward may be to develop a method to rear our lunar herd in-situ on the Moon's surface using only a small portion of the available (currently uninhabited) surface area. 

\begin{figure}[h]
    \centering
    \includegraphics[width=\columnwidth]{\projectpath 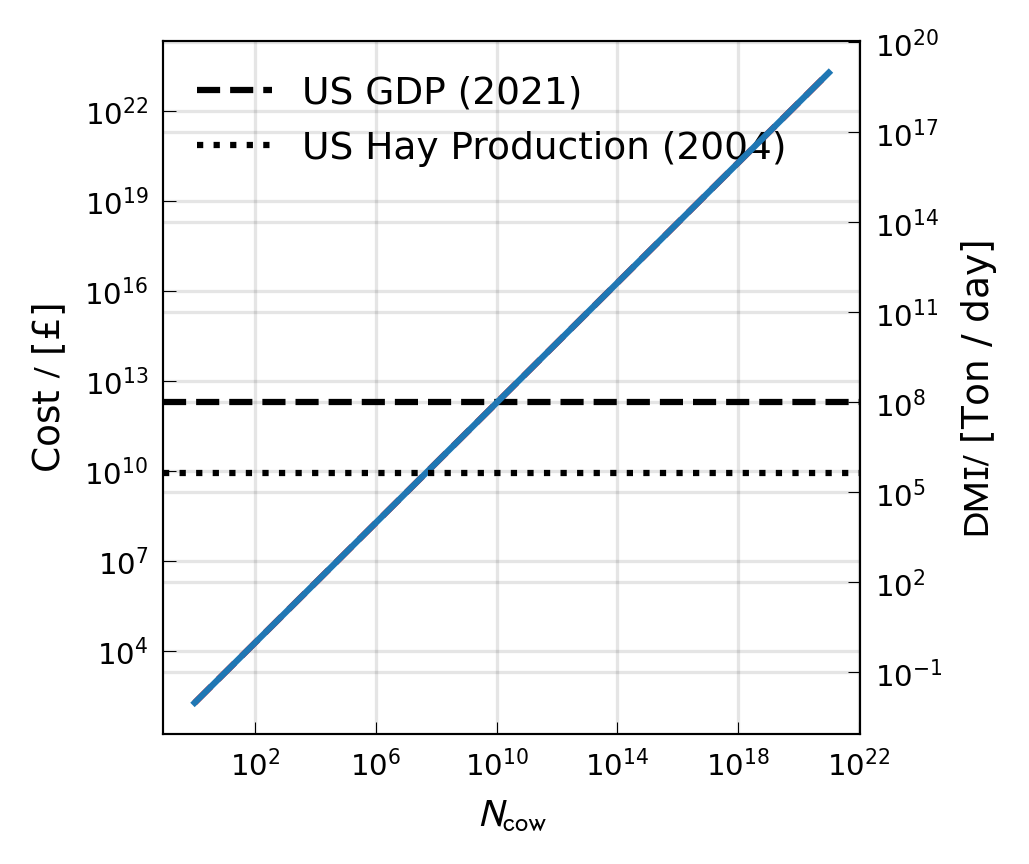}
    \vspace{0.1in}
    \caption{A plot demonstrating the total cost of rearing a planetary herd and the total Dry Matter Index (DMI, a measure of cattle feeding) of the herd per day. For comparison we have included the GDP of the US in 2021 as a dashed line and the total mass of hay produced in a the US in 2004 \cite{hay} (a stand out year) as a dotted line.}
    \label{fig:cost}
    \vspace{0.1in}
\end{figure}

In addition to the area required we must also consider the financial burden and nutritional needs of our herd. In Fig. \ref{fig:cost} we present both of these as a function of the number of cows in the herd. For the total cost of rearing a cow we take an average cost of £190 per cow \cite{cost}. For the required food per day we utilise the Dry Matter Index (DMI), a measure of the dry matter consumed by the cows in Tons per day. As with the area required to rear the cattle, the cost and nutritional requirement of the pure bovine planetoid is so prohibitive as to become impossible. The $\sim10^{19}$ cows would cost at least 2 dex more than the GDP of the US in 2021 and would need to consume per day $\sim10^{17}$ Tons of dry matter. 

For salvation we turn to our lunar herd once more. The $\sim10^6$ cows needed for a steady atmosphere at Earth's orbit would only cost a mere dex less than the US 2021 GDP, a significantly more modest cost! However, we are still left with the prohibitive need to provide this lunar herd nutrition, a requirement which demands a significant portion of the US's entire hay production from 2004 \cite{hay} for a single day. Given the viability of many other factors in seeing to fruition our local lunar herd this is clearly where future efforts should be focused. It is however worth noting, that if we could sustain bovine life at lower temperatures (and relocate the Moon) we would be able to sustain an atmosphere at Jupiter's orbit with a significantly lower strain on rearing area, cost and food production. 

\heading{Conclusion}

In this paper we have investigated the viability of methane atmospheres on bovine planetoids and extended that discussion to a lunar bovine population. We find that a bovine planetoid can sustain a methane based atmosphere at $D>0.1$ AU with $\gtrsim10^{19}$ cows but the cost, necessary rearing area and required nutrition are impossible to meet at the current time. We also find that the temperature of the atmosphere is such that the cows can only survive the temperatures in a small orbital range around the orbit of Earth. However, when considering a lunar herd we not only find that a methane atmosphere can be supported with $\sim10^6$ cows at Earth's orbit but also find that the cost and necessary area are much more manageable. 

Future work should focus on relaxing some of the assumptions made to produce these results while endeavouring to solve some of the problems posed in this paper: How do we provide the necessary nutrition to support the planetary herd? How can we come together as a species to fund this great endeavour? Could this actually be of any use whatsoever?

\heading{Acknowledgements}

We wish to acknowledge Jeff Bezos for his comments on space based factories, which (for reasons that are long lost to the authors) somehow led to a lunch time conversation derailing into the concepts presented in this paper. 

We acknowledge Stephen M. Wilkins for throwing fuel on the flames of this endeavour. We also acknowledge the insightful conversations had with Christopher Brown (Sussex) and Maria Del Carmen Campos Varillas (Sussex) for their contribution to this work.

\heading{Disclaimer}

No cows were harmed in the research presented in this paper, nor do we advocate for any unnecessary harm being brought upon our bovine friends. 

\begingroup
\setlength\bibitemsep{0pt}
\setlength\bibnamesep{0pt}
\printbibliography[heading=subbibliography]
\endgroup

\end{refsection}
\end{document}